\documentclass[a4paper, 10pt]{article}

\usepackage{osameet2}
\usepackage[colorlinks=true,bookmarks=false,citecolor=blue,urlcolor=blue]{hyperref} 
\renewcommand{\thesection}{\Roman{section}}
\renewcommand{\thesubsection}{\thesection.\Roman{subsection}}
\usepackage[utf8]{inputenc}
\usepackage{glossaries}
\usepackage[UKenglish]{babel}
\newcommand{\SetCapsType}{normalcaps}
\makeatletter
\providecommand{\SetCapsType}{smallcaps}
\long\def\@scTrue{smallcaps}
\long\def\@scFalse{normalcaps}
\newcommand{\acroSCaps}[1]{%
 \begingroup
  \ifx\SetCapsType\@scTrue 
    \textsc{#1}%
  \else
    \MakeUppercase{#1}%
  \fi
  \endgroup
}
\makeatother
\newcommand{\nAcronym}[4][]{%
	\newacronym[#1]{#2}{#3}{#4}
}
\makeatletter
\@ifpackageloaded{babel}{
    \newcommand{\usuk}[2]{%
        \iflanguage{USenglish}{#1}{#2}
    }
}{
    \newcommand{\usuk}[2]{%
        #1
    }
}
\makeatother

\usepackage[shortcuts]{extdash}

\nAcronym{2A8PSK}{\acroSCaps{2a8psk}}{2-ary amplitude 8-ary phaseshift keying}
\nAcronym{4D}{\acroSCaps{4d}}{four-dimensional}
\nAcronym{4D64PRS}{\acroSCaps{4d-64prs}}{\usuk{four-dimensional 64-ary polarization-ring-switching}{four-dimensional 64-ary polarisation-ring-switching}}
\nAcronym{5B4D2A8PSK}{\acroSCaps{5b4d-2a8psk}}{5-bit four-dimensional two-amplitude 8-ary phase-shift keying}
\nAcronym{6B4D2A8PSK}{\acroSCaps{6b4d-2a8psk}}{6-bit four-dimensional two-amplitude 8-ary phase-shift keying}
\nAcronym{8D}{\acroSCaps{8d}}{eight-dimensional}\nAcronym{8D2048PRS}{\acroSCaps{8d-2048prs}}{eight-dimensional 2048-ary polarization-ring-switching}
\nAcronym{8D2048PRST1}{\acroSCaps{8d-2048prs-t1}}{eight-dimensional 2048-ary polarization-ring-switching type 1}
\nAcronym{8D2048PRST2}{\acroSCaps{8d-2048prs-t2}}{eight-dimensional 2048-ary polarization-ring-switching type 2}
\nAcronym{ABC}{\acroSCaps{abc}}{automatic bias controller}
\nAcronym{ADC}{\acroSCaps{adc}}{analog-to-digital converter}
\nAcronym{AIR}{\acroSCaps{air}}{achievable information rate}
\nAcronym{AOM}{\acroSCaps{aom}}{acoustic optical modulator}
\nAcronym{AR}{\acroSCaps{ar}}{achievable rate}
\nAcronym{ASE}{\acroSCaps{ase}}{amplified spontaneous emission}
\nAcronym{AWGN}{\acroSCaps{awgn}}{additive white Gaussian noise}
\nAcronym{BER}{\acroSCaps{ber}}{bit error rate}
\nAcronym{BICM}{\acroSCaps{bicm}}{bit-interleaved coded modulation}
\nAcronym{BMD}{\acroSCaps{bmd}}{bit-metric decoding}
\nAcronym{BPD}{\acroSCaps{bpd}}{balanced photo-diode}
\nAcronym{BPS}{\acroSCaps{bps}}{blind phase search}
\nAcronym{BRGC}{\acroSCaps{brgc}}{binary reflected Gray code}
\nAcronym{CCDM}{\acroSCaps{ccdm}}{constant composition distribution matching}
\nAcronym{CD}{\acroSCaps{cd}}{chromatic dispersion}
\nAcronym{CIR}{\acroSCaps{cir}}{channel impulse response}
\nAcronym{ChUT}{\acroSCaps{chut}}{channel under test}
\nAcronym{CUT}{\acroSCaps{cut}}{channel under test}
\nAcronym{CPE}{\acroSCaps{cpe}}{carrier phase estimation}
\nAcronym{CAGR}{\acroSCaps{cagr}}{compound annual growth rate}
\nAcronym{DA}{\acroSCaps{da}}{driver amplifier}
\nAcronym{DAC}{\acroSCaps{dac}}{digital-to-analog converter}
\nAcronym{DCF}{\acroSCaps{dcf}}{\usuk{dispersion compensated fiber}{dispersion compensated fibre}}
\nAcronym{DFB}{\acroSCaps{dfb}}{distributed feedback}
\nAcronym{DGD}{\acroSCaps{dgd}}{differential group delay}
\nAcronym{DM}{\acroSCaps{dm}}{distribution matcher}
\nAcronym{DMGD}{\acroSCaps{dmgd}}{differential mode group delay}
\nAcronym{DP}{\acroSCaps{dp}}{\usuk{dual-polarization}{dual-polarisation}}
\nAcronym{DPC}{\acroSCaps{dpc}}{digital pre-compensation}
\nAcronym{DPE}{\acroSCaps{dpe}}{digital pre-emphasis}
\nAcronym{DPIQ}{\acroSCaps{dp-iqm}}{\usuk{dual-polarization IQ-modulator}{dual-polarisation IQ-modulator}}
\nAcronym{DRE}{\acroSCaps{dre}}{digital resolution enhancer}
\nAcronym{DSO}{\acroSCaps{dso}}{digital sampling oscilloscope}
\nAcronym{DSP}{\acroSCaps{dsp}}{digital signal processing}
\nAcronym{DUT}{\acroSCaps{dut}}{device under test}
\nAcronym{DWDM}{\acroSCaps{dwdm}}{dense wavelength-division multiplexing}
\nAcronym{ECL}{\acroSCaps{ecl}}{external cavity laser}
\nAcronym{ED}{\acroSCaps{ed}}{Eucledian distance}
\nAcronym{EDFA}{\acroSCaps{edfa}}{\usuk{erbium doped fiber amplifier}{erbium doped fibre amplifier}}
\nAcronym{ENOB}{\acroSCaps{enob}}{effective number of bits}
\nAcronym{ESS}{\acroSCaps{ess}}{enumerative sphere shaping}
\nAcronym{FDE}{\acroSCaps{fde}}{\usuk{frequency domain equalizer}{frequency domain equaliser}}
\nAcronym{FEC}{\acroSCaps{fec}}{forward error correction}
\nAcronym{FFT}{\acroSCaps{fft}}{fast Fourier transform}
\nAcronym{FMF}{\acroSCaps{fmf}}{\usuk{few-mode fiber}{few-mode fibre}}
\nAcronym[plural=FM-MCF, firstplural=\usuk{few-mode multi-core fibers}{few-mode multi-core fibres}]{FM-MCF}{\acroSCaps{fm-mcf}}{\usuk{few-mode multi-core fiber}{few-mode multi-core fibre}}
\nAcronym{FWM}{\acroSCaps{fwm}}{four-wave mixing}
\nAcronym{GFF}{\acroSCaps{gff}}{gain flattening filter}
\nAcronym{GMI}{\acroSCaps{gmi}}{\usuk{generalized mutual information}{generalised mutual information}}
\nAcronym{GS}{\acroSCaps{gs}}{geometric shaping}
\nAcronym{GV}{\acroSCaps{gv}}{group-velocity}
\nAcronym{GVD}{\acroSCaps{gvd}}{group-velocity dispersion}
\nAcronym{HDFEC}{\acroSCaps{hd-fec}}{hard decision forward error correction}
\nAcronym[plural=IL, firstplural=insertion losses (\textsc{il})]{IL}{\acroSCaps{il}}{insertion loss}
\nAcronym{IFFT}{\acroSCaps{ifft}}{inverse fast Fourier transform}
\nAcronym{ISI}{\acroSCaps{isi}}{intersymbol interference}
\nAcronym{KK}{\acroSCaps{kk}}{Kramers-Kronig}
\nAcronym{LCOS}{\acroSCaps{LCoS}}{liquid crystal on silicon}
\nAcronym{LDPC}{\acroSCaps{ldpc}}{low-density parity-check}
\nAcronym{LEAF}{\acroSCaps{leaf}}{\usuk{large effective area fiber}{large effective area fibre}}
\nAcronym{LMS}{\acroSCaps{lmf}}{least means square}
\nAcronym{LLR}{\acroSCaps{llr}}{log-likelihood ratio}
\nAcronym{LO}{\acroSCaps{lo}}{local oscillator}
\nAcronym{LP}{\acroSCaps{lp}}{\usuk{linear polarized}{linear polarised}}
\nAcronym{LSPS}{\acroSCaps{lsps}}{\usuk{loop-synchronized polarization scrambler}{loop-synchronised polarisation scrambler}}
\nAcronym{MB}{\acroSCaps{mb}}{Maxwell-Bolzmann}
\nAcronym{MCF}{\acroSCaps{mcf}}{\usuk{multi-core fiber}{multi-core fibre}}
\nAcronym{MDG}{\acroSCaps{mdg}}{mode dependent gain}
\nAcronym[plural=MDL, firstplural=mode dependent losses (\textsc{mdl})]{MDL}{\acroSCaps{mdl}}{mode dependent loss}
\nAcronym{MDM}{\acroSCaps{mdm}}{mode division multiplexing}
\nAcronym{MF}{\acroSCaps{mf}}{matched filter}
\nAcronym{MI}{\acroSCaps{mi}}{mutual information}
\nAcronym{MIMO}{\acroSCaps{mimo}}{multiple-input multiple-output}
\nAcronym{MMA}{\acroSCaps{mma}}{multi-modulus algorithm}
\nAcronym{MMF}{\acroSCaps{mmf}}{\usuk{multi-mode fiber}{multi-mode fibre}}
\nAcronym{MMSE}{\acroSCaps{mmse}}{minimum mean squared error}
\nAcronym{MPLC}{\acroSCaps{mplc}}{multi-plane light converter}
\nAcronym{MSE}{\acroSCaps{mse}}{mean squared error}
\nAcronym{MUX}{\acroSCaps{mux}}{multiplexer}
\nAcronym{MZM}{\acroSCaps{mzm}}{Mach-Zehnder modulator}
\nAcronym{NF}{\acroSCaps{nf}}{noise figure}
\nAcronym{NGMI}{\acroSCaps{ngmi}}{\usuk{normalized generalized mutual information}{normalised generalised mutual information}}
\nAcronym{NIR}{\acroSCaps{nir}}{near infra-red}
\nAcronym{OFDR}{\acroSCaps{ofdr}}{optical frequency-domain reflectometer}
\nAcronym{OMFT}{\acroSCaps{omft}}{optical-multi-format transmitter}
\nAcronym{OSA}{\acroSCaps{osa}}{\usuk{optical spectrum analyzer}{optical spectrum analyser}}
\nAcronym{OSNR}{\acroSCaps{osnr}}{optical signal-to-noise ratio}
\nAcronym{OTDR}{\acroSCaps{otdr}}{optical time-domain reflectometer}
\nAcronym{OTF}{\acroSCaps{otf}}{optical tunable filter}
\nAcronym{OVNA}{\acroSCaps{ovna}}{\usuk{optical vector network analyzer}{optical vector network analyser}}
\nAcronym{PAM}{\acroSCaps{pam}}{pulse-amplitude modulation}
\nAcronym{PAS}{\acroSCaps{pas}}{probabilistic amplitude shaping}
\nAcronym{PAPR}{\acroSCaps{papr}}{peak-to-avarage power ratio}
\nAcronym{PBS}{\acroSCaps{pbs}}{polarization beam splitter}
\nAcronym{PCVD}{\acroSCaps{pcvd}}{plasma chemical vapor depostion}
\nAcronym{PDL}{\acroSCaps{pdl}}{polarization dependent loss}
\nAcronym{PL}{\acroSCaps{pl}}{photonic lantern}
\nAcronym{PMBPSK}{\acroSCaps{pm-bpsk}}{polarization-multiplexed binary phase-shift-keying}
\nAcronym{PMQPSK}{\acroSCaps{pm-qpsk}}{polarization-multiplexed quaternary phase-shift-keying}
\nAcronym{PM8QAM}{\acroSCaps{pm-8qam}}{polarization-multiplexed 8-ary quadrature amplitude modulation}
\nAcronym{PMD}{\acroSCaps{pmd}}{polarization mode dispersion}
\nAcronym{PNOB}{\acroSCaps{pnob}}{physical number of bits}
\nAcronym{PRBS}{\acroSCaps{prbs}}{pseudorandom bit sequence}
\nAcronym{PS}{\acroSCaps{ps}}{probabilistic shaping}
\nAcronym{QAM}{\acroSCaps{qam}}{quadrature amplitude modulation}
\nAcronym{QPSK}{\acroSCaps{qpsk}}{quadrature phase shift keying}
\nAcronym{RRC}{\acroSCaps{rrc}}{root-raised-cosine}
\nAcronym{SamPerSym}{\acroSCaps{sps}}{samples per symbol}
\nAcronym{SDFEC}{\acroSCaps{sd-fec}}{soft decision forward error correction}
\nAcronym{SDM}{\acroSCaps{sdm}}{space-division multiplexing}
\nAcronym{SE}{\acroSCaps{se}}{spectral efficiency}
\nAcronym{SER}{\acroSCaps{ser}}{symbol error rate}
\nAcronym{SA}{\acroSCaps{sa}}{simulated annealing}
\nAcronym{SLM}{\acroSCaps{slm}}{spatial light modulator}
\nAcronym{SSFM}{\acroSCaps{ssfm}}{split-step Fourier method}
\nAcronym{SMF}{\acroSCaps{smf}}{\usuk{single-mode fiber}{single-mode fibre}}
\nAcronym{SNR}{\acroSCaps{snr}}{signal-to-noise ratio}
\nAcronym[firstplural=\usuk{states of polarization (\acroSCaps{sop})}{states of polarisation (\acroSCaps{sop})}]{SOP}{\acroSCaps{sop}}{\usuk{state of polarization}{state of polarisation}}
\nAcronym{SPM}{\acroSCaps{spm}}{self-phase modulation}
\nAcronym{SPS}{\acroSCaps{sps}}{\usuk{synchronized polarization scrambler}{synchronised polarisation scrambler}}
\nAcronym{SSMF}{\acroSCaps{ssmf}}{\usuk{standard single-mode fiber}{standard single-mode fibre}}
\nAcronym{SVD}{\acroSCaps{svd}}{singular value decomposition}
\nAcronym{SWI}{\acroSCaps{swi}}{swept wavelength interferometry}
\nAcronym{TDE}{\acroSCaps{tde}}{\usuk{time domain equalizer}{time domain equaliser}}
\nAcronym{TDMSDM}{\acroSCaps{tdm-sdm}}{time-domain multiplexed space-division multiplexing}
\nAcronym{TE}{\acroSCaps{TE}}{transverse electric}
\nAcronym{TM}{\acroSCaps{TM}}{transverse magnetic}
\nAcronym{TH4D}{\acroSCaps{th-4d}}{time domain hybrid four-dimensional}
\nAcronym{TH4D2A8PSK}{\acroSCaps{th-4d-2a8psk}}{time domain hybrid four-dimensional two-amplitude eight-phase shift keying}
\nAcronym{VOA}{\acroSCaps{voa}}{variable optical attenuator}
\nAcronym{WDM}{\acroSCaps{wdm}}{wavelength division multiplexing}
\nAcronym{WSS}{\acroSCaps{wss}}{wavelength selective switch}
\nAcronym{XPM}{\acroSCaps{xpm}}{cross-phase modulation}
\nAcronym{3DWG}{\acroSCaps{3dwg}}{3D-waveguide}
\usepackage[capitalise]{cleveref}
\usepackage{adjustbox}
\usepackage{bm}
\usepackage{xcolor}
\usepackage[per-mode=symbol]{siunitx}
\usepackage[font=footnotesize]{subcaption}
\renewcommand\thesection{\arabic{section}}
\renewcommand\thesubsection{\thesection.\alph{subsection}}
\usepackage{wrapfig}
\def\airn{\text{AIR}_{N}}
\def\rloss{R_{\text{loss}}}
\usepackage{booktabs}
\usepackage{floatrow}
\newfloatcommand{capbtabbox}{table}[][\FBwidth]
\usepackage{amsmath,amssymb}
\usepackage{newtxmath}
\usepackage{tikz}
\usepackage{pgfplots}
    \pgfplotsset{compat=1.15}
    \usepgfplotslibrary{fillbetween}
    \usepackage{pgfplotstable}
    \usepackage{xstring}
\usetikzlibrary{shapes,arrows}
\usetikzlibrary{calc,positioning}
\definecolor{Cblue}{RGB}{31 119 180}
\definecolor{Cred}{RGB}{214 39 40}
\definecolor{Corange}{RGB}{255 127 14}
\definecolor{Cgreen}{RGB}{44 160 44}
\definecolor{Cpurple}{RGB}{148 103 198}
\definecolor{Cbrown}{RGB}{140 86 75}
\definecolor{Cpink}{RGB}{227 119 194}
\definecolor{Ccyan}{RGB}{23 190 207}
\definecolor{Cgrey}{RGB}{127 127 127}
\definecolor{Cyellow}{RGB}{188 189 34}
\pgfplotsset{compat=1.15}
\pgfmathsetmacro{\spanMult}{75}
\pgfmathsetmacro{\dualPolMult}{2}
\pgfplotsset{Uniform/.style={solid,color=Cblue,thick,mark=triangle*,mark options={solid,fill=Cblue}}}
\pgfplotsset{CCDM-200/.style={solid,color=Cred,thick,mark=*,mark options={solid,fill=Cred}}}
\pgfplotsset{CCDM-3600/.style={dashed,color=Cred,thick,mark=*,mark options={solid,fill=white}}}
\pgfplotsset{ESS-200/.style={solid,color=Cgreen,thick,mark=square*,mark options={solid,fill=Cgreen}}}
\pgfplotsset{FECline/.style={solid,color=black,thick,mark=none,dashed}}
\pgfplotsset{grid style={dashed,lightgray!75}, legend cell align={left}}

\begin{document}
\title{On Optimum Enumerative Sphere Shaping Blocklength at Different Symbol Rates for the Nonlinear Fiber Channel}

\author{
Yunus Can Gültekin\textsuperscript{1},
Olga Vassilieva\textsuperscript{2},
Inwoong Kim\textsuperscript{2},
Paparao Palacharla\textsuperscript{2},
Chigo Okonkwo\textsuperscript{1},
and Alex Alvarado\textsuperscript{1}
}

\address{\textsuperscript{1}{Department of Electrical Engineering, Eindhoven University of Technology, The Netherlands} \\
\textsuperscript{2}{Fujitsu Network Communications Inc., Richardson, 75082 TX, USA} \\
y.c.g.gultekin@tue.nl}

\vspace{-5mm}

\begin{abstract}
We show that a 0.9 dB SNR improvement can be obtained via short-blocklength enumerative sphere shaping for single-span transmission at 56 GBd. 
This gain vanishes for higher symbol rates and a larger number of spans.
\end{abstract}

\maketitle

\vspace{-4mm}

\section{Introduction}

Probabilistic shaping (PS) is a powerful technique that provides signal-to-noise ratio (SNR) gains more than 1 dB for large quadrature amplitude modulation (QAM) constellations and enables rate adaptivity with a fixed constellation and a forward error correction (FEC) code rate~\cite{Fehenberger2016,Buchali2016}.
The former translates into a reach increase~\cite{Amari2019} while thanks to the latter, the available spectrum can be efficiently utilized by dynamically adapting transmission rates according to link qualities.

The most widely used coded modulation framework in which PS is realized is probabilistic amplitude shaping (PAS)~\cite{Bocherer2015}.
PAS combines an amplitude shaper with a systematic FEC code.
Popular amplitude shaping techniques include constant composition distribution matching (CCDM)~\cite{Schulte2016}, enumerative sphere shaping (ESS)~\cite{Gultekin2020,Goossens2019}, etc.
It was observed in~\cite[Fig. 8]{Amari2019} for both CCDM and ESS that when a finite-length amplitude shaper is realized, the resulting SNR decreases for increasing blocklength $N$.
This effect is typically explained via the differences in the temporal structures of the amplitude sequences for different values of $N$ for CCDM~\cite{Fehenberger2020,Wu2021}.
It is possible to deduct from multiple studies scattered around the literature that this dependence of SNR on the shaping blocklength is stronger for smaller symbol rates and shorter distances~\cite{Amari2019,Fehenberger2020,Fehenberger2020_2}.

In this paper, we study the dependency of SNR and achievable information rate (AIR) on ESS blocklength for different symbol rates and link distances for pilot-aided signaling.
We demonstrate that increasing the symbol rate or the number of spans have a similar effect on the characteristics of the nonlinear interference noise (NLIN): both make NLIN behave more like circularly symmetric additive white Gaussian noise (csAWGN).
As the NLIN becomes csAWGN, the dependency of SNR on ESS blocklength weakens.
Finally, we study the relation between SNR and AIR in the linear and nonlinear regimes.
We show that in cases NLIN is csAWGN, there is an almost one-to-one relation between SNR and AIR.

\section{Probabilistic Amplitude Shaping: System Model and Achievable Information Rate}
\begin{figure}[ht]
    \makebox[\textwidth]{\makebox[1\textwidth]{%
    \centering
    \begin{subfigure}[t]{.45\textwidth}%
        \centering 
        \resizebox{\textwidth}{!}{\includegraphics{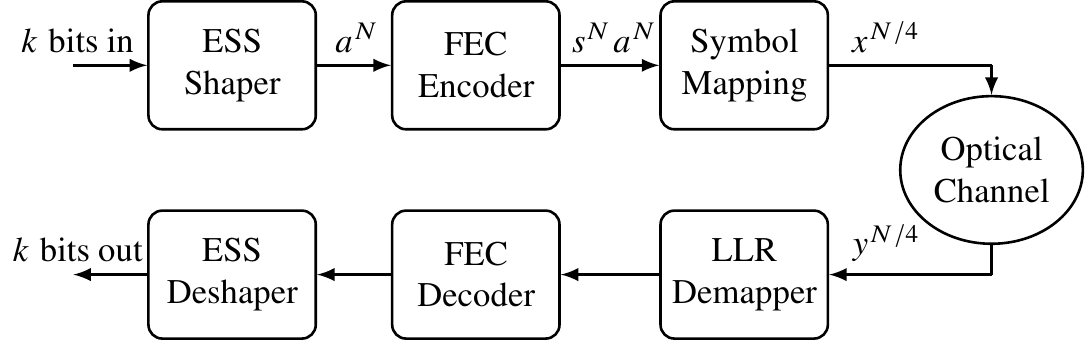}}
        \caption{}%
        \label{fig:PAS}%
    \end{subfigure}%
    \begin{subfigure}[t]{.55\textwidth}%
        \centering 
        \resizebox{\textwidth}{!}{
\begin{tikzpicture}[
line width=0.75pt,
block/.style={rectangle, rounded corners,draw,inner sep=2pt,minimum width=5mm, minimum height=9ex,font=\normalsize,align=center},
block_ell/.style={ellipse, rounded corners,draw,inner sep=2pt,minimum width=5mm, minimum height=10mm,font=\normalsize,align=center},
line/.style = {draw,->,font=\normalsize}
]

\pgfmathsetmacro{\dist}{5}
\pgfdeclareshape{debugpoint}{
    \inheritsavedanchors[from=circle]
    \anchor{center}{\centerpoint}
    \anchorborder{\centerpoint}
    \backgroundpath{
        \pgfpathcircle{\centerpoint}{\radius}
    }
}
\tikzstyle{point-visible} = [debugpoint,inner sep=0pt, minimum size = 2,fill=red] 
\tikzstyle{point-invisible} = [coordinate]
\tikzstyle{point} = [point-invisible] 
\node [rectangle,draw,align=center] (tl) at (0,0) {\includegraphics[trim={2cm 2cm 2cm 2cm}, clip, width=0.5\textwidth]{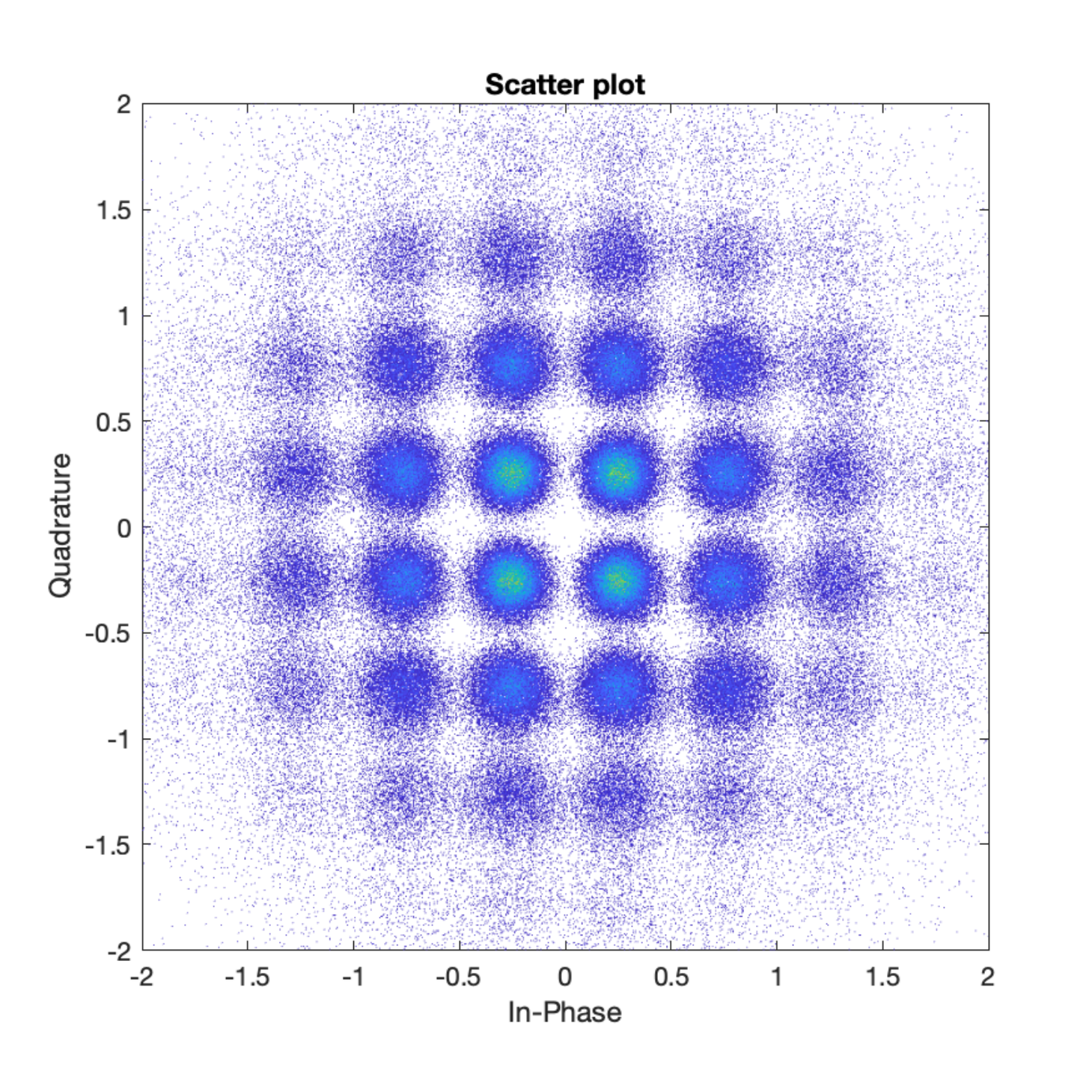}};
\node [rectangle,align=center,draw,minimum width=2.2cm] (tlu) at (0,-0.65*\dist) {\Large {\bf $56$ GBd, $-11.5$ dBm}};
\node [rectangle,align=center,draw,minimum width=2.2cm] (tlu) at (0.5*\dist,0.65*\dist) {\Large {\bf $1$ span, $N=4096$}};
\node [rectangle,draw,align=center] (ml) at (\dist,0) {\includegraphics[trim={2cm 2cm 2cm 2cm}, clip, width=0.5\textwidth]{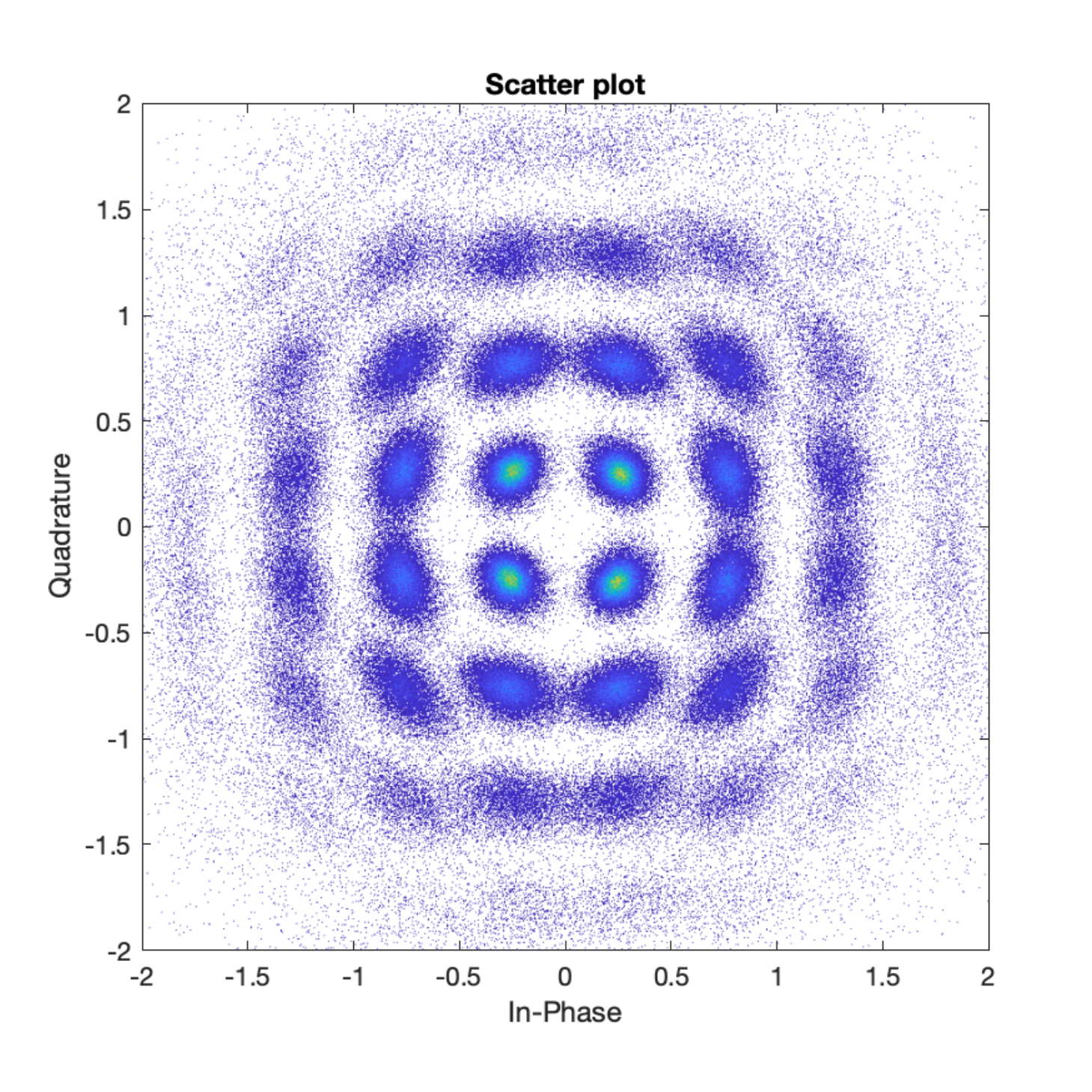}};
\node [rectangle,align=center,draw,minimum width=2.2cm] (tlu) at (\dist,-0.65*\dist) {\Large {\bf $56$ GBd, $9.25$ dBm}};
\node [rectangle,draw,align=center] (mr) at (2*\dist,0) {\includegraphics[trim={2cm 2cm 2cm 2cm}, clip, width=0.5\textwidth]{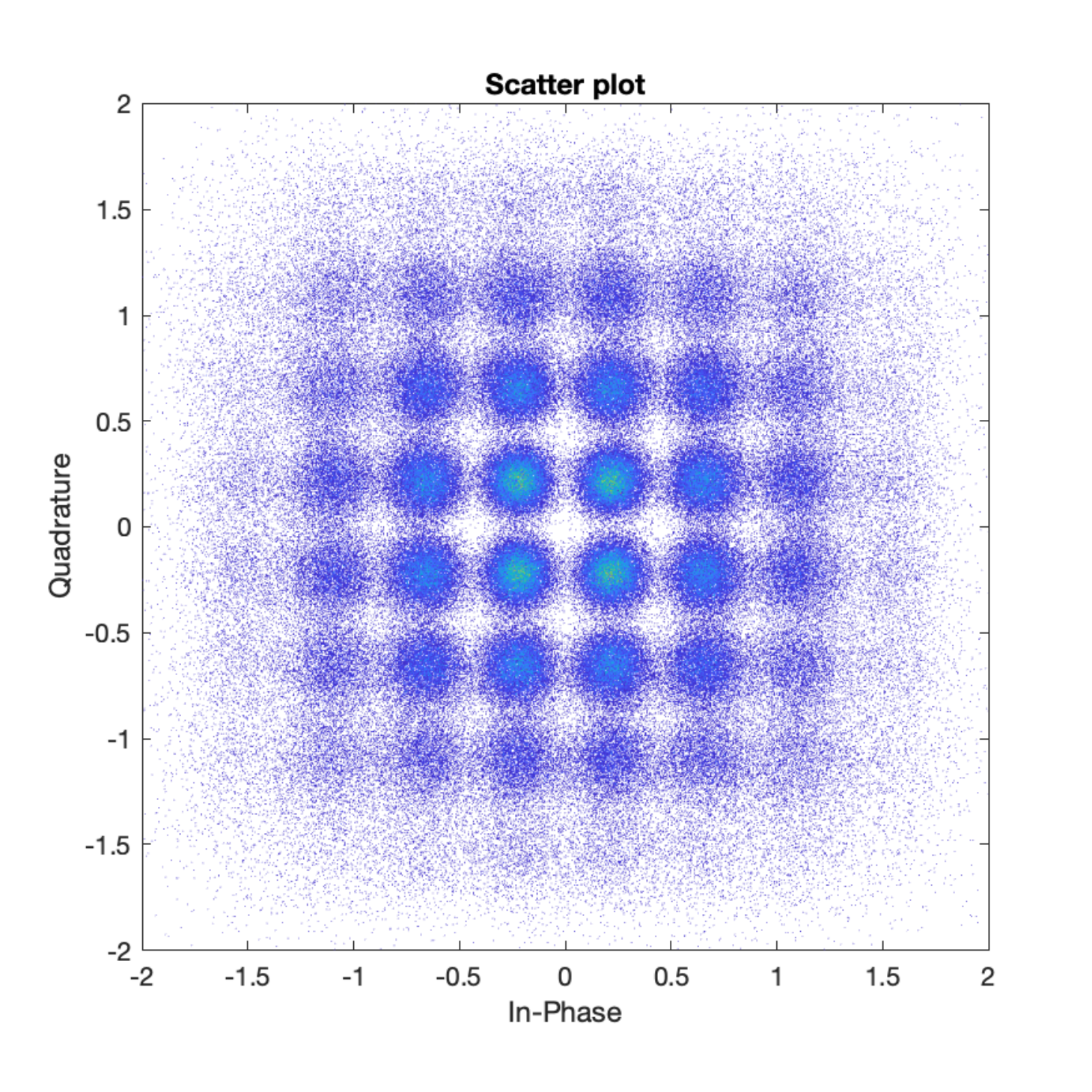}};
\node [rectangle,align=center,draw,minimum width=2.2cm] (tlu) at (2*\dist,-0.65*\dist) {\Large {\bf $160$ GBd, $4$ dBm}};
\node [rectangle,align=center,draw,minimum width=2.2cm] (tlu) at (2.5*\dist,0.65*\dist) {\Large {\bf $10$ span, $N=4$}};
\node [rectangle,draw,align=center] (tr) at (3*\dist,0) {\includegraphics[trim={2cm 2cm 2cm 2cm}, clip, width=0.5\textwidth]{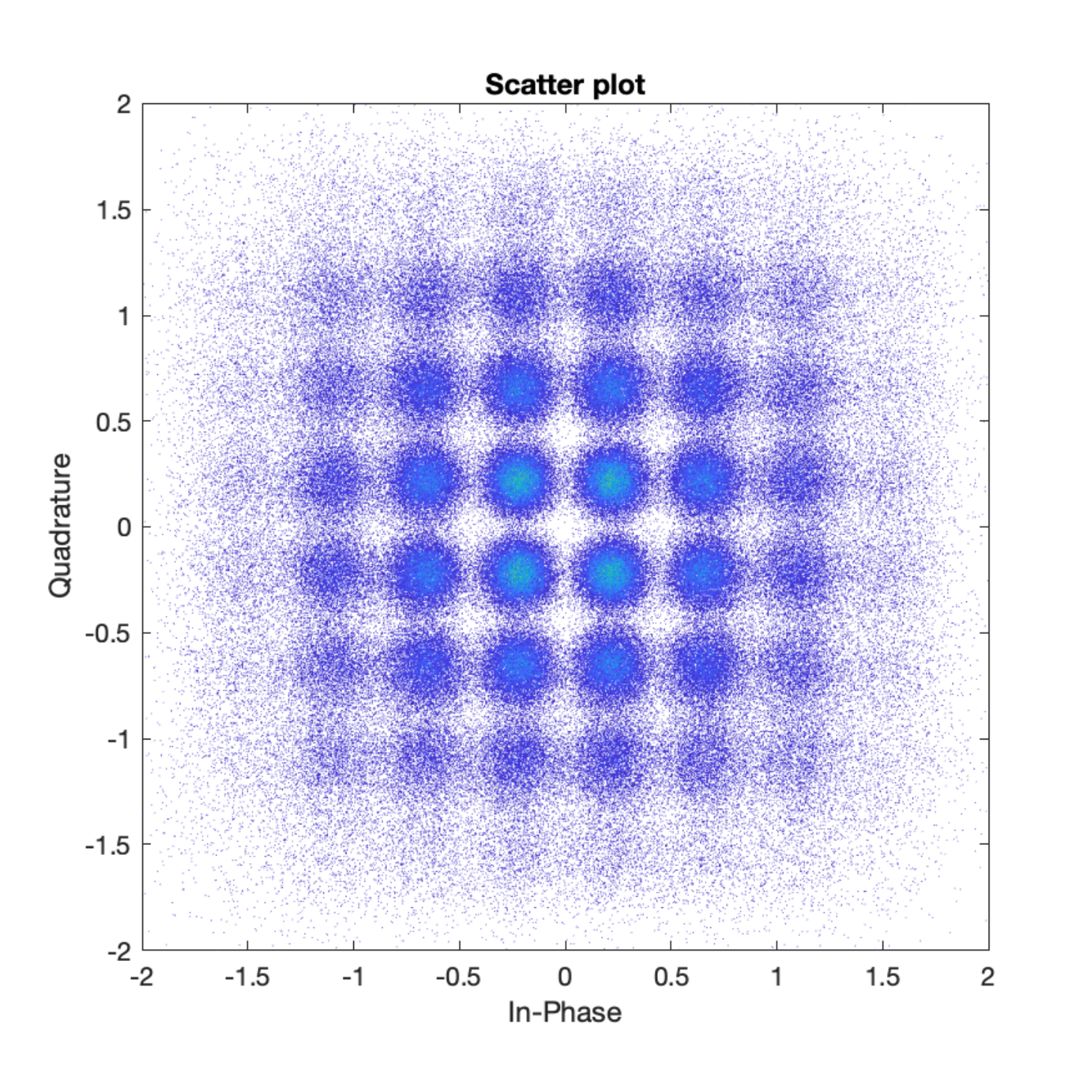}};
\node [rectangle,align=center,draw,minimum width=2.2cm] (tlu) at (3*\dist,-0.65*\dist) {\Large {\bf $160$ GBd, $6$ dBm}};
\end{tikzpicture}
        }
        \caption{}%
        \label{fig:RXconsts}%
    \end{subfigure}
    }}
    \vspace{-5mm}
    \caption{(a) PAS block diagram. (b) Received constellations, all roughly having 16.1 dB SNR.}
\end{figure}
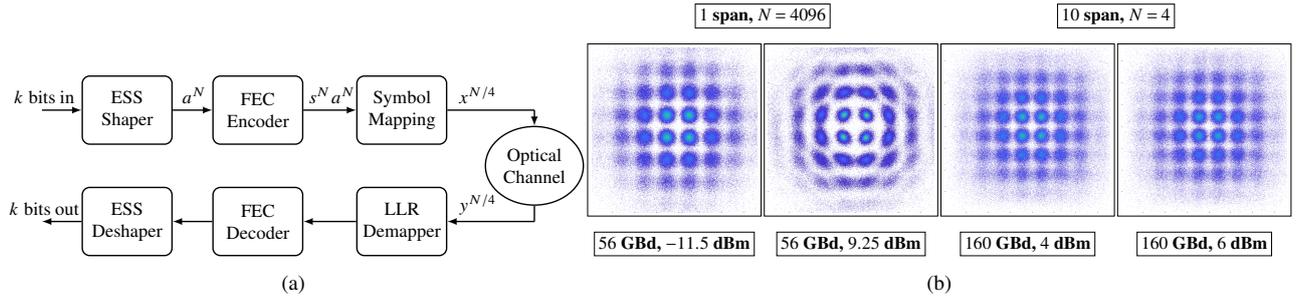

We consider the PAS scheme with ESS as the amplitude shaping technique as shown in Fig.~\ref{fig:PAS}.
First, ESS maps a $k$-bit uniform information string to a shaped $N$-amplitude sequence $a^N$ using the $M$-amplitude-shift keying ($M$-ASK) amplitude alphabet $\{1, 3,\dotsc, M-1\}$.
Then the corresponding binary reflected Gray code (BRGC) is used to convert the amplitudes into amplitude bits.
The following systematic FEC encoder accepts these amplitude bits as the input, preserves their structure, and generates parity bits.
These parity bits are used to determine the signs $s^N$ and hence, an $N$-sequence of $M$-ASK symbols $s^Na^N$ is generated.
Then, a symbol mapper combines these ASK symbols into 4D channel inputs $x^{N/4}$ assuming dual-polarized transmission.
At the receiver, log-likelihood ratios (LLRs) are computed from the received 4D symbols $y^{N/4}$ and passed to the FEC decoder.
Finally, the amplitude deshaper recovers the information bits from the estimated amplitudes.
For this coded modulation scheme, an AIR is given by the finite-length bit-metric decoding rate~\cite{Fehenberger2016}
\begin{equation}\label{rbmd}
     \airn = H(X) - \sum_{i=1}^{m} H(B_i \mid Y)  - \rloss,
\end{equation}
where the rate loss of a finite-length amplitude shaper is defined as $\rloss = H(A)-k/N$.
Here, $H(\cdot)$ is the entropy in bits.
In \eqref{rbmd}, $B_1, B_2,\dotsc, B_m$ represent the binary labels of ASK symbols where $m=\log_2 M$.

There are two aspects of PAS which are not of concern for uniform signaling schemes.
First, the shaping blocklength $N$ plays an important role in the resulting performance.
While the rate loss $\rloss$ decreases for increasing $N$~\cite{Gultekin2020}, the effective SNR tends to decrease~\cite{Amari2019,Fehenberger2020}.
This results in a finite optimum shaping blocklength $N$ at which $\airn$ is maximized~\cite[Fig. 3]{Fehenberger2020}.
This optimum value depends on the parameters of the considered setup, e.g., number of channels, symbol rate, link distance, etc., and can vary from a few dozens of amplitudes to a few hundreds~\cite{Gultekin2021}.
Second, the way ASK symbols are combined into 4D symbols also influences the effective SNR.
There are three trivial ways~\cite[Fig. 3]{Skvortcov2021_HCSS}: In 1D mapping, 4D symbols get their amplitudes from 4 independent shaped sequences, i.e., shaping per real dimension.
In 2D mapping, each polarization gets their amplitudes from 2 independent shaped sequences, i.e., shaping per polarization.
In 4D mapping, amplitudes of a 4D symbol come from the same shaped sequence, i.e., shaping per channel.
Different mapping strategies lead to different temporal structures and probability distributions in channel input sequences, especially for short blocklengths~\cite{Fehenberger2020_2}.
It is observed for CCDM and HCSS that 4D mapping provides the largest SNR~\cite{Fehenberger2020_2,Skvortcov2021_HCSS}.

\section{Numerical Results}

We study the optimum ESS blocklength by simulating the transmission over the nonlinear fiber channel through which the propagation is modeled by the NL Schr\"{o}dinger equation. 
PAS with ESS is realized with shaping blocklengths $N = 4, 8, 16,\dotsc, 4096$ for 64-QAM.
The shaping rate is $k/N=1.5$~bit/1D.
When combined with a rate 5/6 FEC code, this leads to a net information rate of $8$~bit/4D.
However, we do not realize FEC encoding and decoding, and select the signs of the amplitudes at uniformly random.

We simulate transmission over single and multiple spans of fiber using split-step Fourier method. 
Each span is of 80 km length with an attenuation of 0.19 dB/km, a dispersion of 17 ps/nm/km, and a nonlinear parameter of 1.3 1/W/km, followed by an erbium-doped fiber amplifier with a noise figure of 5.5 dB. 
The number of wavelength-division multiplexing (WDM) channels is $11$.
The transmitter generates a dual-polarized signal with a root-raised-cosine pulse with 10\% roll-off.
Both 56 and 160 GBd symbol rates are considered with the channel spacing of 62.5 and 175 GHz, resp..
At the receiver, first the chromatic dispersion, then the constant phase rotation induced by the nonlinearity are compensated.
For the latter, a quadrature phase shift keying pilot is transmitted for every 32 data symbols.
The effective SNR and the $\airn$ in \eqref{rbmd} are averaged over both polarization and 11 channels.

\begin{figure}[ht]
    \makebox[\textwidth]{\makebox[1\textwidth]{%
    \centering
    \begin{subfigure}[t]{.32\textwidth}%
        \centering 
                \resizebox{\textwidth}{!}{\includegraphics{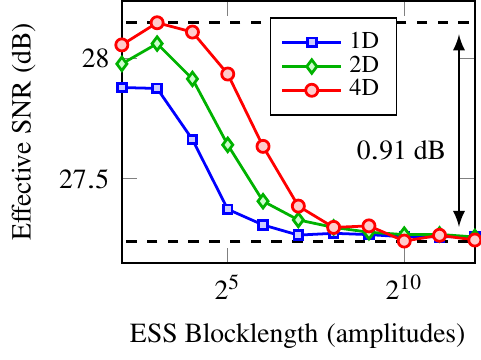}}
        \caption{}
        \label{fig:NvsSNR_1D2D4D}%
    \end{subfigure}
    \begin{subfigure}[t]{.32\textwidth}%
        \centering 
                \resizebox{\textwidth}{!}{\includegraphics{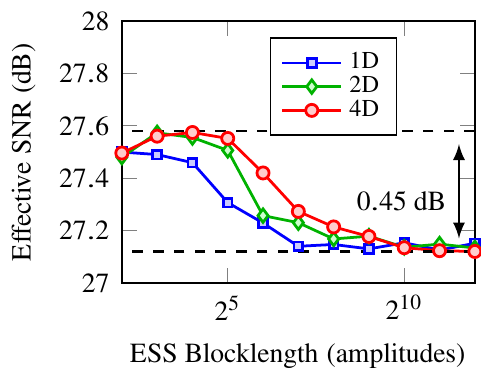}}
        \caption{}%
        \label{fig:NvsSNR_1D2D4D_160G}%
    \end{subfigure}
        \begin{subfigure}[t]{.32\textwidth}%
        \centering 
        \resizebox{\textwidth}{!}{
        \begin{tikzpicture}[font=\tiny]
\usepgfplotslibrary{fillbetween}
\begin{axis}[%
every axis/.append style={font=\footnotesize},
width=0.95\columnwidth,
height=0.78\columnwidth,
xmin=14.2,
xmax=22,
ymin=4.17,
ymax=5,
grid style={dashed,lightgray!75},
xlabel={Effective SNR (dB)},
ylabel={$\airn$ (bit/2D)},
grid=major,
ylabel near ticks,
xlabel near ticks,
legend style={at={(0.65,0.3)},anchor=center,font=\scriptsize,legend cell align=left,row sep=-0.75ex},
]

\addlegendimage{color=black, thick, dashed};
\addlegendentry{AWGN};
\addlegendimage{color=black, thick};
\addlegendentry{160 GBd};
\addlegendimage{color=black, thick, dashed, mark=asterisk, only marks, mark size=1.25pt};
\addlegendentry{56 GBd};

\addplot [name path=block4, color=blue, semithick]
  table[row sep=crcr]{
8.11617337787207	2.55366404567690\\
8.73670102837433	2.72167084997821\\
9.29046647438253	2.88028753731951\\
9.82976354468521	3.03532444047862\\
10.4025986948486	3.19874512579106\\
10.9217707636229	3.35305526166504\\
11.4169228414056	3.50236764019548\\
11.9446567668176	3.65200431645338\\
12.4495329765094	3.79751012235083\\
13.0056768383073	3.95160593574379\\
13.5428762703377	4.09108975405397\\
14.0362503930616	4.21524040870978\\
14.5454573053695	4.33634964705118\\
15.0749571675755	4.44918422982543\\
15.5863498481400	4.54789221449117\\
16.0867893827631	4.63074880320813\\
16.6316659991144	4.71261375151020\\
17.1281973477472	4.77341164664823\\
17.6022960102092	4.82525751042347\\
18.1244765267205	4.86802776907905\\
18.6386450628190	4.90413813089212\\
19.1356644576926	4.93059621044595\\
19.6415314347192	4.95267986606835\\
20.1468493278421	4.96583819977131\\
20.6213885352884	4.97723168404989\\
21.1561279690009	4.98615196522631\\
21.6639465426733	4.99124994209492\\
22.1442193055955	4.99560486976897\\
22.6260075190780	4.99677576262390\\
23.1140131396219	4.99820749721877\\
23.5953658059837	4.99949563471975\\
24.0873549360149	4.99946065661869\\
24.5684517715277	5.00045991137323\\
25.0180503409900	4.99926092422118\\
25.5017445125941	4.99949854651455\\
25.9026550928422	4.99959096802349\\
26.2806395278192	5.00070440014327\\
26.6751643183715	5.00070375311903\\
26.9913924864951	5.00008981989673\\
27.2344039272478	5.00011032650687\\
27.3832000288003	4.99975598750941\\
27.4722510590976	4.99975669296907\\
27.4998895901320	5.00022771415594\\
27.2691713093840	5.00020725505665\\
26.9582763468589	5.00044576019399\\
26.5289590282308	4.99959183552051\\
26.0376361342387	5.00025540422841\\
25.3488697620484	4.99939279276189\\
24.5501176434349	4.99786994882153\\
23.7559148760283	4.99521311144822\\
22.9990990172255	4.99183443403179\\
22.1725316999040	4.98362725312505\\
21.1292168552337	4.96178824582418\\
20.1164269271449	4.92917309321962\\
19.2826710070131	4.89024726001027\\
18.2586946997358	4.81591960608843\\
17.1112916034420	4.69870897624070\\
16.1890142313413	4.58164818289609\\
15.2649951304922	4.42125368025473\\
14.2431271255595	4.21382201785704\\
};
\addplot [name path=block2048, color=red, semithick]
  table[row sep=crcr]{
7.56266325023148	2.78183872544386\\
8.13867519200505	2.93947605530827\\
8.73244436626880	3.10436332016668\\
9.30276787896118	3.26550059189561\\
9.83777650855694	3.41943288910928\\
10.3976979152907	3.57722690104434\\
10.9120272143554	3.72410381095868\\
11.4248524510610	3.87033362539840\\
11.9571469720712	4.01349919431928\\
12.4729242243760	4.14112373375173\\
13.0130858385241	4.26997960882582\\
13.5227325867302	4.38456416395444\\
14.0449137668972	4.48705096388066\\
14.5659244428607	4.58051819828802\\
15.0674603408981	4.65860869587975\\
15.5570161405277	4.72582233956078\\
16.0949753250008	4.78783035842485\\
16.6019039264445	4.83579965406031\\
17.1009445510813	4.87588328489002\\
17.6049333822837	4.90856733467748\\
18.1241201182273	4.93407691562654\\
18.6246778551839	4.95323140278491\\
19.1177089996761	4.96672616590862\\
19.6320410040320	4.97758194696906\\
20.1269286183560	4.98468952441683\\
20.6585250865825	4.99090596904152\\
21.1386809445377	4.99424055707458\\
21.6211379838556	4.99688355661021\\
22.1225658658332	4.99817802285328\\
22.6252384972509	4.99903302434450\\
23.1068971282604	4.99957290585166\\
23.5953655837230	4.99977779228248\\
24.0818992920738	4.99976854896888\\
24.5469881647505	4.99995400919545\\
24.9914084644369	4.99997165137312\\
25.4617305090395	4.99998321761230\\
25.8282667934635	4.99998271045681\\
26.2198887017770	4.99995851635507\\
26.5772985682730	4.99999900429805\\
26.8308072319650	4.99998430452670\\
27.0220770370716	5.00000705379935\\
27.1482915689702	5.00002529271374\\
27.1516259966884	5.00005524037706\\
26.9987968677937	4.99994979714666\\
26.7261178621615	4.99966293208262\\
26.3389797671574	4.99982711544064\\
25.8093182852309	4.99934933805804\\
25.2497195309912	4.99880739901481\\
24.5707793351790	4.99743309283823\\
23.7295226873004	4.99452988778040\\
22.8539458442390	4.98639288855345\\
22.0543422509079	4.97749823443032\\
21.0702740777131	4.95914576701055\\
20.1601334762943	4.93643682364572\\
19.2592794043856	4.90318443362120\\
18.1744093388576	4.84283693405205\\
17.2375094739610	4.77504326631501\\
16.1524313190600	4.67161920524761\\
15.1249940329353	4.54602664113310\\
14.1633353043290	4.40270490805729\\
13.0254211636216	4.19265807557336\\
};
\addplot [color=blue, semithick, dashed, mark=asterisk, only marks, mark size=1.25pt,mark options={fill=blue,solid}]
  table[row sep=crcr]{
12.5213856042672	3.81769768540999\\
13.0519099640281	3.96321496874503\\
13.5920017348651	4.10478859202583\\
14.0935446909816	4.23077465324581\\
14.6418599865027	4.35645629764665\\
15.1212436159296	4.45908287383687\\
15.6620169453493	4.55926051105648\\
16.1436600165627	4.63946852902840\\
16.6689422346082	4.71746124637489\\
17.1577195590190	4.77621227312774\\
17.6835711979139	4.83233149967671\\
18.1756056881204	4.87378389862813\\
18.6794508200712	4.90613100256412\\
19.1817995877556	4.93326543312588\\
19.6900373275636	4.95393921070959\\
20.1986566772171	4.96904005222753\\
20.7042200774462	4.97818093104277\\
21.1969374632632	4.98633289158827\\
21.7281734048285	4.99082465849997\\
22.1930038910365	4.99629525393375\\
22.7057376369906	4.99694175698973\\
23.1917733994632	4.99809459552063\\
23.6849401039188	4.99855438814859\\
24.1592565281816	5.00060964462236\\
24.6606652988397	4.99888218775202\\
25.1220011179062	4.99921400816688\\
25.5897073232827	5.00005443562081\\
26.0030239499772	5.00006998431570\\
26.4490703642630	5.00059837632998\\
26.8293594090653	4.99988265971888\\
27.1929349779360	4.99990341091775\\
27.4827172395277	5.00034402173269\\
27.6968829960776	4.99992793381271\\
27.8470733416153	5.00064948988753\\
27.8790469845068	4.99932221730878\\
27.7847570318309	4.99954822470115\\
27.6080318155232	4.99962004837954\\
27.2377719146544	5.00067881842424\\
26.7785638740696	5.00051918150417\\
26.2310212205674	4.99964743457221\\
25.5526971897300	4.99875961786054\\
24.7972220220950	4.99584575282765\\
23.9959055661200	4.99226337739055\\
23.1293286250597	4.98261388350927\\
22.1991838114124	4.96607915932968\\
21.2727318292054	4.93979895942072\\
20.3043877013562	4.90045307179483\\
19.3548011268834	4.84326479221528\\
18.3151940114021	4.76421178103275\\
17.3294794691937	4.66183136836597\\
16.3495822936585	4.53512112400496\\
15.3709736098041	4.38085248982135\\
14.3343093596939	4.17809977829844\\
13.2994838090348	3.93744413125774\\
12.2331716841597	3.65867370502360\\
11.2282224977877	3.37148632699097\\
10.1407123137714	3.04968046728545\\
9.08403769004753	2.73386524218625\\
7.98736301561453	2.41011652182818\\
6.81459996707194	2.07404490075612\\
5.64507631287012	1.75540572204575\\
};
\addplot [color=blue, semithick, dashed, mark=asterisk, no marks, mark size=1.25pt,mark options={fill=blue,solid}]
  table[row sep=crcr]{
12	3.63934747231152\\
12.2000000000000	3.69917855871140\\
12.4000000000000	3.75868009435888\\
12.6000000000000	3.81776410283218\\
12.8000000000000	3.87633619771862\\
13	3.93429619086366\\
13.2000000000000	3.99153887585807\\
13.4000000000000	4.04795497108880\\
13.6000000000000	4.10343220162489\\
13.8000000000000	4.15785649559944\\
14	4.21111326864639\\
14.2000000000000	4.26308876926072\\
14.4000000000000	4.31367145843452\\
14.6000000000000	4.36275339824371\\
14.8000000000000	4.41023162584706\\
15	4.45600949124773\\
15.2000000000000	4.49999793886600\\
15.4000000000000	4.54211671428931\\
15.6000000000000	4.58229547843046\\
15.8000000000000	4.62047481178548\\
16	4.65660709168799\\
16.2000000000000	4.69065722562081\\
16.4000000000000	4.72260322401833\\
16.6000000000000	4.75243659680861\\
16.8000000000000	4.78016255941608\\
17	4.80580003620980\\
17.2000000000000	4.82938145252439\\
17.4000000000000	4.85095231039182\\
17.6000000000000	4.87057054794207\\
17.8000000000000	4.88830568791865\\
18	4.90423778671918\\
18.2000000000000	4.91845620157755\\
18.4000000000000	4.93105819967298\\
18.6000000000000	4.94214743877745\\
18.8000000000000	4.95183235423071\\
19	4.96022449123250\\
19.2000000000000	4.96743682439091\\
19.4000000000000	4.97358210789272\\
19.6000000000000	4.97877129939260\\
19.8000000000000	4.98311209863311\\
20	4.98670763790654\\
20.2000000000000	4.98965535584721\\
20.4000000000000	4.99204607892777\\
20.6000000000000	4.99396332674377\\
20.8000000000000	4.99548284814405\\
21	4.99667238599480\\
21.2000000000000	4.99759165938692\\
21.4000000000000	4.99829254394884\\
21.6000000000000	4.99881942408999\\
21.8000000000000	4.99920968587732\\
22	4.99949431609909\\
22.2000000000000	4.99969857202331\\
22.4000000000000	4.99984268735754\\
22.6000000000000	4.99994258275569\\
22.8000000000000	5.00001055355797\\
23	5.00005591283420\\
};
\addplot [color=red, semithick, dashed, mark=asterisk, only marks, mark size=1.25pt,mark options={fill=red,solid}]
  table[row sep=crcr]{
12.5533813607984	4.16168219083993\\
13.0634449069806	4.28273779260047\\
13.5715173523617	4.39079064785130\\
14.1087099600781	4.50055524780785\\
14.6511136608718	4.59601945244931\\
15.1367805028452	4.67040347417265\\
15.6469358631646	4.73801574836297\\
16.1574279882268	4.79611176671699\\
16.6655993239656	4.84425073222432\\
17.1826593176705	4.88329415309300\\
17.6482693252966	4.90947480055405\\
18.1863427641937	4.93644077455288\\
18.6924814269963	4.95579363828050\\
19.1910961895948	4.96904290156063\\
19.6907305665175	4.97874875432396\\
20.1970878880618	4.98645317187654\\
20.6915047360928	4.99163605876650\\
21.1955065090673	4.99490285505267\\
21.6871396107200	4.99705679710729\\
22.1984500229567	4.99815797581172\\
22.6756717585058	4.99895036452269\\
23.1748622445855	4.99967635070712\\
23.6513765401679	4.99972512348396\\
24.1524958391121	4.99983751751689\\
24.6229563294941	4.99997380516385\\
25.0679445432275	4.99996145059713\\
25.4903657066721	5.00002259477981\\
25.9250048018883	5.00001195337268\\
26.3124103039252	5.00000961589177\\
26.6375309129065	5.00002250073699\\
26.9482495930687	4.99999082153766\\
27.1032237802326	5.00000832747811\\
27.2605767672105	5.00003300763410\\
27.2432467078614	4.99999470816803\\
27.1378323848989	4.99993402994883\\
26.9380996258556	4.99983237474324\\
26.5241848579804	4.99931286090076\\
26.1097163572141	4.99909529933115\\
25.4731144099460	4.99693478080811\\
24.7664632931574	4.99268130824121\\
24.0279336441897	4.98683002132071\\
23.2195796209761	4.97512005319959\\
22.3294153310229	4.95653368570780\\
21.4214987572045	4.92874628730431\\
20.4670893514344	4.89142258661488\\
19.5030387462516	4.84635965609683\\
18.5504021385878	4.78641725242904\\
17.5774360145469	4.70974308511871\\
16.6103478615057	4.62165701173438\\
15.5423223412130	4.50110309789539\\
14.5486950524922	4.36747521762356\\
13.5888603191437	4.21135172242333\\
12.5278672932848	4.00341401650073\\
11.4771078313637	3.75668493203280\\
10.4667086222701	3.49129087198170\\
9.45586806319353	3.20029514962308\\
8.35760883711095	2.87392739767855\\
7.10226240965476	NaN\\
6.18194001937107	2.23675790815828\\
4.96118656406158	NaN\\
3.71730230899335	NaN\\
};
\addplot [color=red, semithick, dashed, mark=asterisk, no marks, mark size=1.25pt,mark options={fill=red,solid}]
  table[row sep=crcr]{
12	4.03118108685473\\
12.2000000000000	4.08637177418678\\
12.4000000000000	4.14059339141798\\
12.6000000000000	4.19373170897786\\
12.8000000000000	4.24567277082566\\
13	4.29630421646091\\
13.2000000000000	4.34551662601690\\
13.4000000000000	4.39320486538309\\
13.6000000000000	4.43926940978837\\
13.8000000000000	4.48361762571352\\
14	4.52616499217927\\
14.2000000000000	4.56683624326960\\
14.4000000000000	4.60556641420049\\
14.6000000000000	4.64230177342765\\
14.8000000000000	4.67700062338879\\
15	4.70963395270292\\
15.2000000000000	4.74018592324160\\
15.4000000000000	4.76865417664701\\
15.6000000000000	4.79504994676507\\
15.8000000000000	4.81939796718455\\
16	4.84173616667068\\
16.2000000000000	4.86211514969923\\
16.4000000000000	4.88059746445767\\
16.6000000000000	4.89725666639639\\
16.8000000000000	4.91217619149704\\
17	4.92544805961090\\
17.2000000000000	4.93717143422093\\
17.4000000000000	4.94745107050230\\
17.6000000000000	4.95639568826520\\
17.8000000000000	4.96411630998267\\
18	4.97072460633514\\
18.2000000000000	4.97633129234315\\
18.4000000000000	4.98104461604168\\
18.6000000000000	4.98496897872131\\
18.8000000000000	4.98820372106222\\
19	4.99084210317517\\
19.2000000000000	4.99297049890869\\
19.4000000000000	4.99466781616217\\
19.6000000000000	4.99600514581877\\
19.8000000000000	4.99704563279872\\
20	4.99784455416698\\
20.2000000000000	4.99844958172494\\
20.4000000000000	4.99890120051287\\
20.6000000000000	4.99923325048794\\
20.8000000000000	4.99947355651901\\
21	4.99964461179054\\
21.2000000000000	4.99976428160483\\
21.4000000000000	4.99984649814993\\
21.6000000000000	4.99990192165218\\
21.8000000000000	4.99993854899511\\
22	4.99996225685318\\
22.2000000000000	4.99997727218573\\
22.4000000000000	4.99998656815316\\
22.6000000000000	4.99999218785480\\
22.8000000000000	4.99999550156189\\
23	4.99999740528232\\
};
\path[name path=axis] (axis cs:-10,0) -- (axis cs:-10,10);
\addplot[blue, opacity=0.2] fill between[of=block4 and axis];
\addplot[red, opacity=0.2] fill between[of=block2048 and axis];
\addplot [color=black, semithick, densely dotted]
  table[row sep=crcr]{
16.15	4\\
16.15	6\\
};        
\draw[-latex, thin] (15.35, 4.7)--(15.1, 4.85) node[anchor=west, pos=1, rotate=30] {};
\draw[] (15.14, 4.88)--(15.14, 4.88) node[rotate=40] {\scalebox{0.9}{$P<P_{\text{opt}}$}};
\draw[-latex, thin] (14.83, 4.5)--(15.15, 4.26) node[anchor=west, pos=1, rotate=30] {};
\draw[] (15.74, 4.24)--(15.74, 4.24) node[rotate=20] {\scalebox{0.9}{$P>P_{\text{opt}}$}};
\end{axis}
\end{tikzpicture}
        }
        \caption{}
        \label{fig:EyeFig}
    \end{subfigure}
    }}
    \vspace{-5mm}
    \caption{{\bf 1-span, 11-chan.,} ESS-$N$: (a) $N$ vs. SNR at 56 GBd. (b) $N$ vs. SNR at 160 GBd. (c) SNR vs. $\airn$ for $N=4$ (blue) and $N=4096$ (red).}
    \label{fig:fig2}
\end{figure}
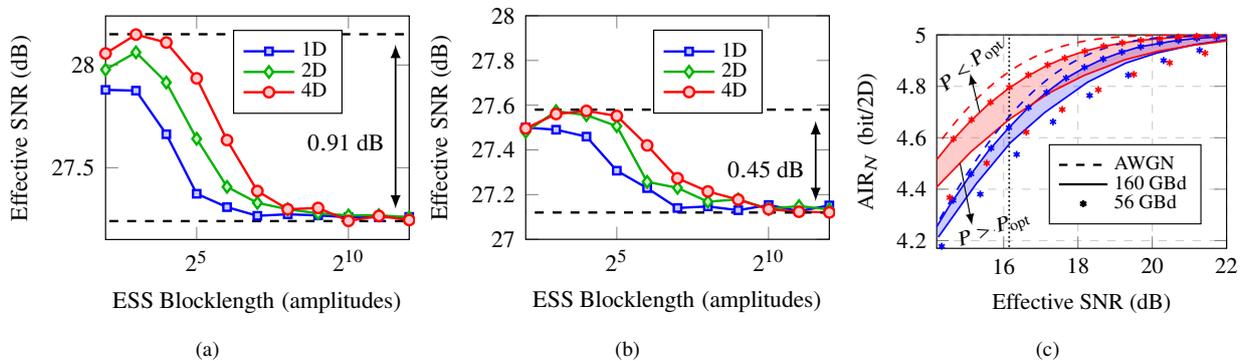

In Figs.~\ref{fig:NvsSNR_1D2D4D} and~\ref{fig:NvsSNR_1D2D4D_160G}, effective SNR at the optimum launch power $P_{\text{opt}}$ is shown as a function of ESS blocklength $N$ for 1D, 2D, and 4D mapping strategies at 56 and 160 GBd, resp.
The first observation is that 4D mapping generally leads to higher values of SNR.
Though, we also remark that if the effective SNR was plotted against the blocklength in 4D symbols (which would require dividing the values on the x-axis by 2 and 4 for 2D and 4D mapping, resp.), the SNR behavior would seem roughly the same for all three mapping strategies.
Nevertheless, the largest SNR in both Figs.~\ref{fig:NvsSNR_1D2D4D} and~\ref{fig:NvsSNR_1D2D4D_160G} is achieved for 4D mapping.
Consequently, we used 4D mapping for the simulations presented in the remainder of the paper.
The second observation is that as $N$ increases, the effective SNR decreases until it converges roughly around $N=256$.
The decrease in effective SNR as $N$ increases is larger for 56 GBd than it is for 160 GBd.
For instance the difference in SNR for $N=4$ and $4096$ is 0.91 dB for 56 GBd, while it is 0.45 dB for 160 GBd.
Next, we will provide a possible explanation for this via the properties of the NLIN.

\begin{figure}[ht]
    \makebox[\textwidth]{\makebox[1\textwidth]{
    \centering
    \begin{subfigure}[t]{.32\linewidth}
        \centering 
        \resizebox{\textwidth}{!}{
        \begin{tikzpicture}[font=\tiny]
\usepgfplotslibrary{fillbetween}
\begin{axis}[
every axis/.append style={font=\footnotesize},
width=0.95\columnwidth,
height=0.78\columnwidth,
xmin=13,
xmax=17,
ymax=4.84,
grid style={dashed,lightgray!75},
xlabel={Effective SNR (dB)},
ylabel={$\airn$ (bit/2D)},
grid=major,
ylabel near ticks,
xlabel near ticks,
legend style={at={(0.65,0.3)},anchor=center,font=\scriptsize,legend cell align=left,row sep=-0.75ex},
]
\addlegendimage{color=black, thick, dashed};
\addlegendentry{AWGN};
\addlegendimage{color=black, thick};
\addlegendentry{56 GBd};
\addplot [name path=block4, color=blue, semithick, mark=asterisk, no marks, mark size=1.25pt,mark options={fill=blue,solid}]
  table[row sep=crcr]{
12.7190988160910	3.81511742892783\\
13.2199410870819	3.95432829258896\\
13.7322164473416	4.08670641902073\\
14.1945103213827	4.20437737444166\\
14.6622387313638	4.31140132812204\\
15.1362007824985	4.41337765552323\\
15.5646592707794	4.49963419294009\\
15.9582580699405	4.57315605429098\\
16.2955666161477	4.62604703904792\\
16.6336523244334	4.67548418572212\\
16.8574191365472	4.71070361098479\\
16.9810769611180	4.72487764780246\\
17.0040960856133	4.72179822195217\\
16.9331817136955	4.71222424229998\\
16.7624405390839	4.68815622030109\\
16.3883101950131	4.62740938771544\\
15.9816836320066	4.55829172135264\\
15.3376750958702	4.44020372758633\\
14.6248264349947	4.28653350475741\\
13.9588320623790	4.12813512748481\\
13.0147418386153	3.88543677508072\\
12.0956983793137	3.63248186726242\\
11.0760159189174	3.34675836893518\\
9.94181212319334	NaN\\
8.94897167912919	2.72810418462969\\
};
\addplot [color=blue, semithick, dashed, mark=asterisk, no marks, mark size=1.25pt,mark options={fill=blue,solid}]
  table[row sep=crcr]{
13	3.93429619086366\\
13.2000000000000	3.99153887585807\\
13.4000000000000	4.04795497108880\\
13.6000000000000	4.10343220162489\\
13.8000000000000	4.15785649559944\\
14	4.21111326864639\\
14.2000000000000	4.26308876926072\\
14.4000000000000	4.31367145843452\\
14.6000000000000	4.36275339824371\\
14.8000000000000	4.41023162584706\\
15	4.45600949124773\\
15.2000000000000	4.49999793886600\\
15.4000000000000	4.54211671428931\\
15.6000000000000	4.58229547843046\\
15.8000000000000	4.62047481178548\\
16	4.65660709168799\\
16.2000000000000	4.69065722562081\\
16.4000000000000	4.72260322401833\\
16.6000000000000	4.75243659680861\\
16.8000000000000	4.78016255941608\\
17	4.80580003620980\\
};
\addplot [name path=block8192, color=red, semithick, mark=asterisk, no marks, mark size=1.25pt,mark options={fill=red,solid}]
  table[row sep=crcr]{
12.4932689253391	4.14976599568647\\
13.0108319641310	4.27190345439626\\
13.5075506664588	4.37950384018403\\
13.9698163373682	4.47507483127216\\
14.4461149389990	4.56054270174382\\
14.8948095849059	4.63405706050759\\
15.3454932876306	4.69800837672504\\
15.7275891996764	4.74764502056918\\
16.0457733307044	4.78119511202343\\
16.3419908866287	4.81082932989083\\
16.5472024237966	4.82819972764196\\
16.6775306323545	4.83892787253390\\
16.6727815463256	4.83430237496598\\
16.5840245451668	4.82381330746265\\
16.3480297856748	4.79725600495535\\
15.9693157976685	4.75280064300133\\
15.4167488792938	4.67953629411360\\
14.8832763621407	4.60302811365686\\
14.1187724944082	4.47421528351941\\
13.1966666117420	4.28664252072249\\
12.4093776410673	4.10279171670591\\
11.4625097693517	3.86534121445262\\
10.4739046680753	3.59883178220764\\
9.37373357999190	NaN\\
8.18947460359959	NaN\\
};
\addplot [color=red, semithick, dashed, mark=asterisk, no marks, mark size=1.25pt,mark options={fill=red,solid}]
  table[row sep=crcr]{
13	4.29742155074804\\
13.2000000000000	4.34660428392591\\
13.4000000000000	4.39426106577591\\
13.6000000000000	4.44029248928752\\
13.8000000000000	4.48460606155972\\
14	4.52711742369995\\
14.2000000000000	4.56775149164810\\
14.4000000000000	4.60644350023914\\
14.6000000000000	4.64313993300156\\
14.8000000000000	4.67779932028524\\
15	4.71039288854815\\
15.2000000000000	4.74090504423163\\
15.4000000000000	4.76933367682520\\
15.6000000000000	4.79569026763009\\
15.8000000000000	4.81999979347166\\
16	4.84230041822010\\
16.2000000000000	4.86264296941671\\
16.4000000000000	4.88109020246869\\
16.6000000000000	4.89771586060664\\
16.8000000000000	4.91260354488401\\
17	4.92584541468205\\
};
\draw (axis cs:13.5,4) ellipse [x radius=0.05cm, y radius=0.3cm];
\node[align=center,font=\scriptsize] (aa) at (axis cs:13.5,3.85){$N=4$};
\draw (axis cs:13.5, 4.38) ellipse [x radius=0.05cm, y radius=0.4cm];
\node[align=center,font=\scriptsize,rotate=10] (aa) at (axis cs:13.6, 4.62){$N=4096$};
\path[name path=axis] (axis cs:-10,0) -- (axis cs:-10,10);
\addplot[blue, opacity=0.2] fill between[of=block4 and axis];
\addplot[red, opacity=0.2] fill between[of=block8192 and axis];
\end{axis}
\end{tikzpicture}
        }
        \caption{}
        \label{fig:NvsAIRn1020span}
    \end{subfigure}
    \begin{subfigure}[t]{.32\linewidth}
        \centering 
                        \resizebox{\textwidth}{!}{\includegraphics{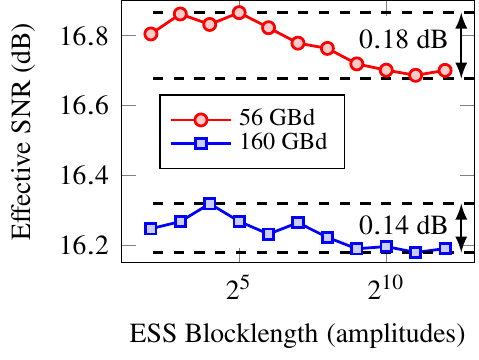}}
        \caption{}
        \label{fig:NvsSNR10span}
    \end{subfigure}
    \begin{subfigure}[t]{.32\linewidth}
        \centering 
                \resizebox{\textwidth}{!}{\includegraphics{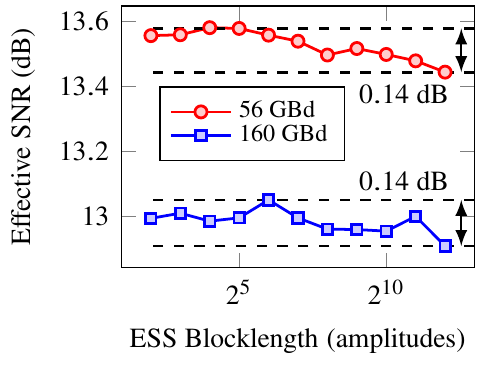}}
        \caption{}
        \label{fig:NvsSNR20span}
    \end{subfigure}
    }}
    \vspace{-5mm}
    \caption{{\bf 11-chan.,} ESS-$N$: (a) SNR vs. $\airn$ after 10 spans. (b) $N$ vs. SNR after 10 spans. (c) $N$ vs. SNR after 20 spans.}
    \label{fig:fig3}
\end{figure}
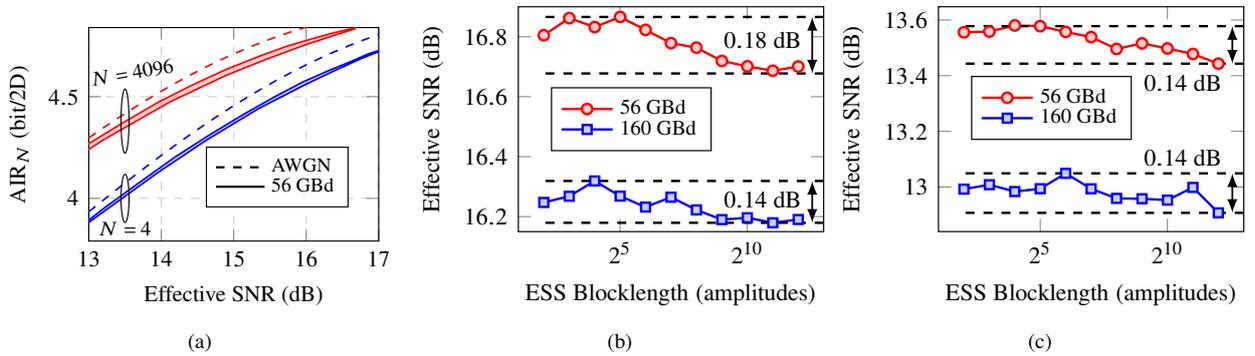

In Fig.~\ref{fig:EyeFig}, we show $\airn$ as a function of the effective SNR.
Since the effective SNR behaves as shown, e.g., in~\cite[Fig. 1]{Dar2015}, meaning that it has a maximum at the optimum launch power $P_{\text{opt}}$, the same effective SNR value is obtained for two launch powers: $P_1<P_{\text{opt}}$ and $P_2>P_{\text{opt}}$.
Thus, SNR vs. $\airn$ curves consist of two ``branches'' with identical SNR ranges as shown in Fig.~\ref{fig:EyeFig}.
The gap between these two branches, i.e., the {\it eye}, can be considered as a very rough indicator of the NLIN characteristics.
As an example, first consider Fig.~\ref{fig:RXconsts} where the received constellations are shown after 1 span for 56 GBd and after 10 spans for 160 GBd at two input powers $P$ each, one smaller than $P_{\text{opt}}$, the other larger.
In each case, an SNR of roughly 16.1 dB (dotted lines in Figs.~\ref{fig:EyeFig} and~\ref{fig:NvsAIRn1020span}) is obtained.
We see that for transmission over 1 span at 56 GBd, while the noise clouds are roughly csAWGN at $P=-11.5$~dBm$<P_{\text{opt}}$, they become more elliptically shaped at $P=9.25$~dBm$>P_{\text{opt}}$ due to nonlinearity, especially for outer constellation points.
On the other hand for transmission over 10 spans at 160 GBd, noise clouds are roughly csAWGN at both input powers.

Now, consider again Fig.~\ref{fig:EyeFig}.
We see that there is a larger eye between the branches of SNR vs. $\airn$ curves for 56 GBd transmission, connected to the losing of circular-symmetry of the noise as input power increases, see Fig.~\ref{fig:RXconsts}.
On the other hand, the eye is narrower for 160 GBd transmission, connected to the circular-symmetry present at both smaller and higher launch powers.
We associate the larger SNR swing observed for 56 Gbd in Figs.~\ref{fig:NvsSNR_1D2D4D} and~\ref{fig:NvsSNR_1D2D4D_160G} with this change in the circular-symmetry of NLIN, i.e., with a wider eye.
We also observe that the eye is narrower for $N=4$ (blue area) than it is for $N=4096$ (red area). 
This also confirms that using shorter ESS blocklengths mitigate NLIN, a fact that can be deducted from Figs.~\ref{fig:NvsSNR_1D2D4D} and~\ref{fig:NvsSNR_1D2D4D_160G}.
In Fig.~\ref{fig:EyeFig}, we also show $\airn$ for the AWGN channel.
We see that a wider eye implies that the $\airn$ gap to the AWGN case is larger, which also implies non-csAWGN behavior.

Finally, consider Fig.~\ref{fig:NvsAIRn1020span} where we show SNR vs. $\airn$ at 56 GBd after 10 spans.
Unlike Fig.~\ref{fig:EyeFig}, the branches are almost on top of each other, i.e., the eye is quite narrow, which is expected since the NLIN behaves as csAWGN in this case as we exhibited in Fig.~\ref{fig:RXconsts}.
Consequently, we expect to see a weaker dependence of effective SNR on $N$.
To confirm this, in Figs.~\ref{fig:NvsSNR10span} and~\ref{fig:NvsSNR20span}, we show effective SNR at the optimum launch power $P_{\text{opt}}$ vs. $N$ for 10-span and 20-span transmission, resp., at 56 and 160 GBd.
In all four cases, the difference between the maximum and the minimum SNR is below 0.2 dB, while it tends to be smaller for the higher symbol rate or the longer link distance.
Thus, we conclude from Figs.~\ref{fig:fig2} and~\ref{fig:fig3} that as the symbol rate or the number of spans increase, the dependence of effective SNR on ESS blocklength $N$ weakens.
This is accompanied with a non-csAWGN NLIN behavior and a visible eye in the SNR vs. $\airn$ curve.

\section{Conclusions}

We have shown via fiber simulations that the effect of ESS blocklength on SNR is only significant for short links and relatively small symbol rates.
This is shown to have connection to the circular-symmetry of the NLI noise:
When the NLI noise becomes circularly symmetric---which is the case for increasing number of spans and symbol rates---, the dependence of SNR on blocklength weakens. 

\vspace{-2mm}

\noindent\footnotesize\emph{\\
The work of Y.C. G\"{u}ltekin and A. Alvarado has received funding from the ERC under the European Union's Horizon 2020 research and innovation programme via the Starting grant FUN-NOTCH (grant ID: 757791) and via the Proof of Concept grant SHY-FEC (grant ID: 963945).}

\end{document}